\documentclass[prb,aps,twocolumn,draft,tightenlines,
showpacs,floatfix]{revtex4}
\usepackage{psfig}
\usepackage{amssymb}
\usepackage{amsmath}
\usepackage{colordvi}

\begin{document}

\title{Spin relaxation in an InAs quantum dot in the presence of
    terahertz driving fields}
\author{J. H. Jiang}
\author{M. W. Wu}
\thanks{Author to whom correspondence should be addressed}
\email{mwwu@ustc.edu.cn.}
\affiliation{Hefei National Laboratory for Physical Sciences at
  Microscale, University of Science and Technology of China, Hefei,
  Anhui, 230026, China}
\affiliation{Department of Physics,
  University of Science and Technology of China, Hefei,
  Anhui, 230026, China}
\altaffiliation{Mailing Address.}

\date{\today}
\begin{abstract}
The spin relaxation in a 1D InAs quantum dot with
the Rashba spin-orbit coupling under driving THz magnetic fields
is investigated by developing a kinetic
equation with the help of the Floquet-Markov theory,
which is  generalized to
the system with the spin-orbit coupling,  to include both
the strong driving field and the electron-phonon scattering.
The spin relaxation time can be effectively prolonged or shortened by the
terahertz magnetic field depending on the frequency and strength
of the terahertz magnetic field.  The effect can be understood as
the sideband-modulated spin-phonon interaction.
This offers an additional way to manipulate the spin relaxation time.
\end{abstract}

\maketitle

\section{Introduction}
One of the goals of semiconductor spintronics\cite{spintronics,wolf}
is to realize quantum information processing based on electron/hole spins.
Coherent oscillations of spin state driven by an AC magnetic/electric field,
which is the key of such a goal, have been broadly
studied.\cite{Gupta,Kato,Rashba1,Cheng,Jiang,Tarucha,Golovach,Duckheim,Koppens}  Experiments have demonstrated successfully the coherent oscillations of
electron spin by optical Stark effect in quantum
wells on femtosecond time scale and by
gigahertz-gate-voltage-controlled $g$-tensor modulation.\cite{Gupta,Kato}
Recently, driven coherent oscillations of single spin in quantum dots (QDs)
at hundreds of MHz has also been realized.\cite{Koppens}
Theoretically, Rashba and Efros showed perturbatively that
electron spin can be manipulated by a weak in-plane time-dependent
electric field via the spin-orbit
coupling (SOC), such as the Rashba\cite{Rashba} and
the Dresselhaus\cite{Dresselhaus} couplings.\cite{Rashba1}
They called it the electric-dipole spin resonance (EDSR).
Similar schemes have also been proposed in QDs.\cite{Tarucha,Golovach}
Cheng and Wu have discussed {\em non}-perturbatively the effect of an intense
terahertz (THz) electric field on  two-dimensional electron gas
(2DEG) with the Rashba SOC where the spin
splitting is of the order of THz.\cite{Cheng}
They showed that a THz electric field can strongly affect the density
of states of the electron system and induce a THz spin oscillations.
Similar effects have also been studied in QD by
Jiang {\em et al.}\cite{Jiang}
However, up till now there is no study on the spin dissipation effect,
{\em i.e.}, spin dephasing/relaxation in the AC-field driven systems,
especially from a fully microscopic approach.
Recently Duckheim and Loss have studied the EDSR in
the presence of disorder in a 2DEG,
where the dissipation effect is introduced by a relaxation time.\cite{Duckheim}
It has been shown in the system without the SOC that
the dissipation under driving field can be very different from
the one without the driving field.\cite{DrivenReview1,DrivenReview2}
A full microscopic kinetic equation under driving field
can be developed with the help of Floquet-Markov
theory.\cite{Kohler} The Floquet-Markov theory combines  the
Floquet theory which can solve the time-dependent (periodic)
Sch\"odinger equation {\em non-perturbatively} with
the Born-Markov approximation which is frequently used in deriving the
 kinetic equation with dissipations.

In the present paper, we extend the Floquet-Markov approach to the
system with the SOC. The change in density of states by the THz laser
field\cite{Cheng,Jiang}  implies a modification
of the spin related scattering and then the spin relaxation time.
We study the spin relaxation in an InAs QD by developing a
Floquet-Markov type kinetic equation with
the electron-acoustic-phonon scattering.
The spin relaxation time is obtained by numerically solving the
kinetic equation.
We develop the model and formulism in Sec.\ II and present the numerical
results in Sec.\ III. We conclude in Sec.\ IV.

\section{Model and formalism}
 We construct our theory in a 1D
QD in  which electron is strongly confined in the $x$-$y$
directions and relatively weaker along the $z$-axis.  This kind of
QDs have been realized experimentally and are attracting more and
more interests due to the good controllability of the size, shape,
position, and electronic structures.\cite{1DQD,1DQD2} A static
magnetic field and a THz electric/magnetic field, which  can be
provided by the free electron laser,\cite{Ramian} are applied
along the $z$-axis. The total Hamiltonian is then given by
\begin{equation}
  H_{tot} = H_e + H_p + H_{ep},
  \label{H_tot}
\end{equation}
with $H_p = \sum_{{\bf q}\eta} \hbar \omega_{{\bf q}\eta} a_{{\bf
    q}\eta}^{\dagger} a_{{\bf q}\eta} $ and $H_{ep}= \sum_{{\bf q}\eta}M_{{\bf q}\eta}
(a_{{\bf q}\eta}+a_{{\bf -q}\eta}^{\dagger})\exp(i{\bf q} \cdot {\bf
  r})$ representing the phonon and the electron-phonon
interaction Hamiltonian respectively.
The electron Hamiltonian in the Coulomb gauge reads\cite{Winkler}
\begin{equation}
  H_e = \frac{{\mathbf P}^2}{2 m^{\ast}} + V_c({\mathbf r}) +
  H_{so}({\mathbf P}) + H_Z\ ,
  \label{H_e}
\end{equation}
in which ${\mathbf P} = -i\hbar \mbox{\boldmath$\nabla$\unboldmath} +
e/c{\mathbf A}(t)$ with $\mathbf{A}(t)=-c\mathbf{E_{1}}\cos(\Omega t)/\Omega
+ \frac{1}{2}[\mathbf{B}_0+\mathbf{B}_1\cos(\Omega t)]\times\mathbf{r}$
denoting the vector potential induced by the THz electric/magnetic
field and the static magnetic field.
$m^{\ast}$ denotes the electron effective mass.
$V_c({\mathbf r})$ is the confining potential of the QD which is
taken to be square well with infinite well depth in each direction.\cite{1DQD2}
$H_{so}(\mathbf P)$ is the SOC term, which consists of the
Rashba term due to the structure inversion asymmetry\cite{Rashba}
and the Dresselhaus term due to the bulk inversion
asymmetry.\cite{Dresselhaus} In InAs the Rashba SOC is the
dominant contribution. $H_Z = \frac{1}{2} g \mu_B
(\mathbf{B}_0+\mathbf{B}_1\cos(\Omega t))\cdot
\mbox{\boldmath$\sigma$\unboldmath}$ represents the
Zeeman term.
When the transverse confinement is strong enough, only the lowest
subband is considered.
Then the Hamiltonian of the electron can be reduced to an
effective Hamiltonian
$H_{eff} = \frac{P_z^2}{2 m^{\ast}} + V(z)+
  H_{so}^{R}(P_z) + H_Z$,
where $P_z = p_z - eE_{1}\cos(\Omega t)/\Omega$ and
$p_z=-i\hbar\frac{\partial}{\partial z}$.
The Rashba term is
$H_{so}^{R} = \frac{e\gamma_{R}}{\hbar} (\sigma_x E_y - \sigma_y
  E_x) P_z$ with $E_x$ and $E_y$ being the electric fields along the $x$-
and $y$-directions which are due to the structure inversion asymmetry and can
be controlled by the gate voltage.\cite{Grundler} For simplicity
we choose $E_y = 0$ in our investigation and $H_{so}^{R}= - e\frac{\gamma_{R}}{\hbar} E_x
\sigma_yP_z \equiv\frac{\alpha_R}{\hbar}\sigma_yP_z$.  This
choice will not change the results of the calculation as
one can always take a unitary
transformation in the spin space to transfer any other configuration into our
simplified one.
The effective Hamiltonian is then written as
\begin{equation}
H_{eff}(t) = H_0 + H_1(t) + H_2(t)\ ,
\label{H_eff}
\end{equation}
with
\begin{eqnarray}
&&\hspace{-0.6cm}H_0 = \frac{p_z^2}{2m^{\ast}} + V(z) + \frac{\alpha_R}{\hbar}
  \sigma_y p_z +  \frac{1}{2}g\mu_B B_0 \sigma_z\ ,\\
&&\hspace{-0.6cm}H_1(t) = [\frac{1}{2}g\mu_B B_1\sigma_z-\frac{eE_{1}}{\Omega}
(\frac{p_z}{m^{\ast}}+\frac{\alpha_R}{\hbar}\sigma_y)]\cos(\Omega t)\ ,\\
&&\hspace{-0.6cm}H_2(t) = \frac{e^2E_{1}^2\cos^2(\Omega t)}{2m^{\ast}\Omega^2}\ .
\end{eqnarray}
Hence $H_{eff}(t+T_{ac}) = H_{eff}(t)$ and $T_{ac}=\frac{2\pi}{\Omega}$.
We observe that $H_2(t)$ is only a function of time, and does not
  contain any other physical variables of electron. Thus it only induces
a universal phase and has no contribution to the kinetics of
  the system.
The corresponding Schr\"odinger equation
\begin{equation}
  i\hbar\frac{\partial}{\partial t}\Psi(t)=H_{eff}(t)\Psi(t)
\label{SchroeEq}
\end{equation}
can be solved via the Floquet-Fourier approach
developed by Shirley\cite{Shirley}
and applied lately in systems with the SOC by Cheng and Wu.\cite{Cheng}
The solution is
\begin{equation}
  \Psi_{\lambda}(z, t) = e^{-i\varepsilon_{\lambda} t}
  \sum_{n=-\infty}^{\infty}\sum_{\alpha}F_{n\alpha}^{\lambda}
  \phi_{\alpha}(z) e^{in\Omega t}\ ,
\label{psi}
\end{equation}
in which $\{\phi_{\alpha}(z)\}$ is a complete set of the wave
functions, chosen here to be the eigen-functions of a
infinite-depth-square-well potential $V(z)$.\cite{Cheng,Jiang}
$\{\varepsilon_{\lambda}\}$ and
$\{F_{n\alpha}^{\lambda}\}$ are the
quasi-energies and the eigen-vectors of the following equations:\cite{Shirley}
\begin{equation}
  \sum_{m=-\infty}^{\infty}\sum_{\beta} \langle\alpha
  n|{\cal{H}_F}|\beta m\rangle  F_{m\beta}^{\lambda} =
  \varepsilon_{\lambda} F_{n\alpha}^{\lambda}\ ,
  \label{eigen}
\end{equation}
where  $\langle{\bf r},t|\alpha n\rangle\equiv\phi_{\alpha}({\bf r})
e^{in\Omega t}$, ${\cal{H}_F} = H_{eff}(t) - i\partial_t$ and $\langle\alpha
n|{\cal{H}_F}|\beta m\rangle = H_{\alpha \beta}^{n-m} + m\Omega \delta_{\alpha
  \beta}\delta_{n m}$ with $H^n = \frac{1}{T} \int_{0}^{T}\!\!d{t}
e^{-in\Omega t} H_{eff}(t)$ representing the $n$-th Fourier component of the
effective Hamiltonian. Due to the periodicity of ${\cal{H}_F}$, the eigenvalues
are also periodic and can be written as
$\varepsilon_{\lambda,l}=\varepsilon_{\lambda,0} +l\Omega$ where
$\varepsilon_{\lambda,0}$ is the eigenvalue in the region
$(-\Omega/2,\Omega/2]$.
It is  noted that $\varepsilon_{\lambda,l}$ and
$\varepsilon_{\lambda,0}$ correspond to the same physical solution
to the Schr\"odinger equation. In the following we denote
$\varepsilon_{\lambda,0}$ with $\varepsilon_{\lambda}$ for simplicity.
The eigenvectors of the
eigen-equations satisfy the orthogonal and complete relations
\begin{eqnarray}
&&\sum_{n,\alpha} F_{n\alpha}^{\lambda_1,l_1\ast}
  F_{n\alpha}^{\lambda_2,l_2} = \delta_{\lambda_1\lambda_2}\delta_{l_1l_2}\ ,
\\
&&\sum_{\lambda,l} F_{n_1\alpha_1}^{\lambda,l\ast}
  F_{n_2\alpha_2}^{\lambda,l} = \delta_{\alpha_1\alpha_2}\delta_{n_1n_2}\ .
\end{eqnarray}
From these relations one obtains the orthogonal and complete relations of the
Floquet wavefunctions\cite{DrivenReview1,Shirley}
\begin{eqnarray}
  \sum_{\lambda} \Psi^{\ast}_{\lambda}(z,t) \Psi_{\lambda}(z^{\prime},t)
  = \delta(z - z^{\prime}), \\
  \int_{-\infty}^{\infty}\!\!d{z} \Psi^{\ast}_{\lambda}(z,t)
  \Psi_{\lambda^{\prime}}(z,t) = \delta_{\lambda,\lambda^{\prime}}\ .
\end{eqnarray}
The wavefunction [Eq.\ (\ref{psi})] includes two significant effects,
one is the sideband effect\cite{sideband} and the other
is the AC Stark effect.\cite{ACStark} The former refers to the
many  frequencies $\varepsilon_{\lambda}-n\Omega$ in the wavefunction and
the later represents the field-induced change of  $\varepsilon_{\lambda}$.

The Floquet wavefunctions which contain all the dynamic properties of
the electron system without the electron-phonon coupling,
give an optimal base to solve the equation of motion of the reduced
density matrix of the electron system. The Floquet-Markov method
which combines the Floquet solution of the electron Hamiltonian and
the Born-Markov approach to solve the equation of motion under strong
AC driving field was developed by Kohler {\sl et al.} in the absence of the
SOC.\cite{Kohler} Generally this method
works well when the AC driven electron system is in the dynamic stable
regime and the system interacts weakly with a Markovian reservoir with a
damping rate much less than any eigen-frequency of the system.
The latter requirement can be satisfied for almost every case in the spin
decoherence problem, as  spins are generally expected to have a very long
coherence time.\cite{LongSDT} In addition, studies have shown that, as
well as keeping good quantitative results, the Floquet-Markov method
has the advantage of being easy to handle numerically compared with
the rather complicated path-integral approach.\cite{Kohler}
Thus this method is very useful in the study of
relaxation/dephasing in nano-structures.
In the present paper, we apply this method to
the systems with the SOC to study the spin relaxation in
QD due to the electron-acoustic-phonon scattering under the
THz driving field.

With the standard Feynman-Vernon initial condition
$\rho(t_0)=\rho^e(t_0)\otimes \rho_{eq}^{p}$, where
$\rho$ is the density matrix of the electron and phonon system;
$\rho^e$ is the
density matrix of the electron system and $\rho_{eq}^{p}$ represents
the density matrix of the equilibrium phonon reservoir,
 and within the Born-Markov
approximation, the reduced density matrix of the electron system
satisfies the following equation:
\begin{widetext}
\begin{eqnarray}
\frac{\partial}{\partial t} \rho^e = - \frac{i}{\hbar} [H_{e}(t),
  \rho^e]
%\nonumber\\
%&&\hspace{-1.cm}
- \frac{1}{\hbar^2}
  \int_{0}^{\infty}\!\!d{\tau} \mbox{Tr}_{p}\{[H_{ep},
[\tilde{H}_{ep}(t-\tau,t),
  \rho^e\otimes\rho_{eq}^{p}]]\}
\end{eqnarray}
with $\mbox{Tr}_{p}$ standing for the trace over the phonon degree of freedom.
$\tilde{H}_{ep}(t-\tau,t)=U_{0}^{\dagger}(t-\tau,t)H_{ep}U_{0}(t-\tau,t)$,
in which $ U_{0}(t-\tau,t) = {\cal P}_t
\exp[-\frac{i}{\hbar}\int_{t}^{t-\tau}\!\!d{t^{\prime}}
{(H_e(t^{\prime}) + H_{p})}]$
with ${\cal P}_t$ denoting the time-ordering operator.
Next we express this integral-differential equation of operators in a
complete base of Floquet wavefunctions denoted by
$\{|\lambda(t)\rangle\}$. Using the complete relation
$\sum_{\lambda}|\lambda(t)\rangle\langle\lambda(t)|=1$,\cite{DrivenReview1,Shirley}
after some simple algebra, one has
\begin{eqnarray}
  \frac{\partial}{\partial t} \rho_{\lambda_1\lambda_2}^e
&=& - \frac{1}{\hbar^2}
  \int_{0}^{\infty}\!\!d{\tau} \sum_{\lambda_3\lambda_4}
\mbox{Tr}_{p}
  (H_{\lambda_1\lambda_3}^{ep}\tilde{H}_{\lambda_3\lambda_4}^{ep}
  \rho_{\lambda_4\lambda_2}^e\otimes\rho_{eq}^{p} -
  \tilde{H}_{\lambda_1\lambda_3}^{ep}\rho_{\lambda_3\lambda_4}^e\otimes
  \rho_{eq}^{p}H_{\lambda_4\lambda_2}^{ep}) + H.c.\nonumber\\
  &=& - \frac{1}{\hbar^2}
  \int_{0}^{\infty}\!\!d{\tau} \sum_{\lambda_3\lambda_4} \sum_{{\bf
      q}\eta} (X_{\lambda_1\lambda_3}^{{\bf q}}
  \tilde{X}_{\lambda_3\lambda_4}^{-{\bf q}}
  \rho_{\lambda_4\lambda_2}^e -
  \tilde{X}_{\lambda_1\lambda_3}^{-{\bf q}}
  \rho_{\lambda_3\lambda_4}^eX_{\lambda_4\lambda_2}^{{\bf
    q}})\langle\langle A_{{\bf q}\eta}(\tau)A_{-{\bf
      q}\eta}\rangle\rangle + H.c.
\label{rhoe}
\end{eqnarray}
where $X_{\lambda_1\lambda_2}^{{\bf q}}= \langle\lambda_1(t)|\exp(i{\bf
  q}\cdot{\bf r})|\lambda_2(t)\rangle$, $A_{{\bf q}\eta}(t) =M_{{\bf q}\eta}(a_{-{\bf
    q}\eta}^{\dagger}e^{i\omega_{{\bf q}\eta}t}+a_{{\bf
    q}\eta}e^{-i\omega_{{\bf q}\eta}t})$ and
$\tilde{X}_{\lambda_1\lambda_2}^{{\bf
    q}}=\langle\lambda_1(t)|{U_{0}^e}^{\dagger}(t-\tau,t)\exp(i{\bf
  q}\cdot{\bf r})U_{0}^e(t-\tau,t)|\lambda_2(t)\rangle$ with
$U_{0}^e(t-\tau,t)= {\cal P}_t
\exp[-\frac{i}{\hbar}\int_{t}^{t-\tau}\!\!d{t^{\prime}} H_e(t^{\prime})]$.
$\langle\langle\cdots\rangle\rangle$ in Eq.\ (\ref{rhoe})
represents the statistical average over the phonon equilibrium
distribution. By substituting the Floquet wave
functions into $X$, one obtains
$X_{\lambda_1\lambda_2}^{{\bf q}} = \sum_{k}
e^{i\Delta_{\lambda_1\lambda_2k}t}X_{\lambda_1\lambda_2k}^{{\bf q}}$
where
$\Delta_{\lambda_1\lambda_2k}=(\varepsilon_{\lambda_1}
-\varepsilon_{\lambda_2})/\hbar+k\Omega$
and $X_{\lambda_1\lambda_2k}^{{\bf
    q}}=\sum_{n}\sum_{\alpha~\beta}
    F_{n\alpha}^{\lambda_1\ast}F_{n+k~\beta}^{\lambda_2}
    \langle\alpha|\exp(i{\bf q}\cdot{\bf r})|\beta\rangle
    ={X_{\lambda_2\lambda_1-k}^{-{\bf q}\ast}}$.
As the Floquet wavefunctions are the solutions to
Eq.\ (\ref{SchroeEq}), one has
\begin{eqnarray}
 \tilde{X}_{\lambda_1\lambda_2}^{{\bf q}} &=&
 \langle\lambda_1(t)|{U_{0}^e}^{\dagger}(t-\tau,t)\exp(i{\bf q}\cdot{\bf
   r})U_{0}^e(t-\tau,t)|\lambda_2(t)\rangle
= \langle\lambda_1(t-\tau)|\exp(i{\bf q}\cdot{\bf
   r})|\lambda_2(t-\tau)\rangle \nonumber\\
 &=& \sum_{k} e^{i\Delta_{\lambda_1\lambda_2k}(t-\tau)}
X_{\lambda_1\lambda_2k}^{{\bf q}}\ .
\end{eqnarray}
Therefore, Eq.\ (\ref{rhoe}) reads
\begin{eqnarray}
\frac{\partial}{\partial t} \rho_{\lambda_1\lambda_2}^e &=&
  - \frac{1}{\hbar^2} \sum_{\lambda_3\lambda_4}\sum_{k_1 k_2}\sum_{{\bf q}\eta}
  \pi |M_{{\bf q}\eta}|^{2}\{ X_{\lambda_1\lambda_3k_1}^{{\bf
      q}}X_{\lambda_4\lambda_3k_2}^{{\bf
q}\ast}\rho_{\lambda_4\lambda_2}e^{i(\Delta_{\lambda_1\lambda_3k_1}
-\Delta_{\lambda_4\lambda_3k_2})t}
C_{{\bf q}\eta}
(\Delta_{\lambda_4\lambda_3k_2})\nonumber\\
&&\mbox{} -X_{\lambda_4\lambda_2k_1}^{{\bf
      q}}X_{\lambda_3\lambda_1k_2}^{{\bf
        q}\ast}\rho_{\lambda_3\lambda_4}
e^{i(\Delta_{\lambda_4\lambda_2k_1}-\Delta_{\lambda_3\lambda_1k_2})t}
C_{{\bf q}\eta}(\Delta_{\lambda_3\lambda_1k_2})\} + H.c.\ ,
\label{EOM}
\end{eqnarray}
with $C_{{\bf q}\eta}(\Delta)=\bar{n}(\omega_{{\bf
    q}\eta})\delta(\Delta+\omega_{{\bf q}\eta})+(\bar{n}(\omega_{{\bf
    q}\eta})+1)\delta(\Delta-\omega_{{\bf q}\eta})$. Here
$\bar{n}(\omega_{{\bf q}\eta})$ is the Bose distribution function.
The summations over $k_1$ and $k_2$ range from $-\infty$ to
  $\infty$. The terms with $C_{{\bf q}\eta}(\Delta_{\lambda_3\lambda_1k_2})$
  describe the $k_2$-photon-assisted scattering.
These equations are still time-dependent. With the rotating wave
approximation (RWA), one can sweep out the time-dependent terms which
oscillate much faster than the damping rate of the density matrix\cite{Kohler}
and Eq.\ (\ref{EOM}) is further simplified into
\begin{equation}
  \frac{\partial}{\partial t} \rho_{\lambda_1\lambda_2}^e =
  - \sum_{\lambda_3\lambda_4}
  \Lambda_{\lambda_1\lambda_2\lambda_3\lambda_4}
  \rho_{\lambda_3\lambda_4}^e
\label{steady}
\end{equation}
with
\begin{eqnarray}
  \Lambda_{\lambda_1\lambda_2\lambda_3\lambda_4} &=&
  \Big\{ \frac{1}{\hbar^2} \sum_{k_1,k_2}\sum_{{\bf q}\eta}
  \pi |M_{{\bf q}\eta}|^{2}[ \sum_{\lambda_5}X_{\lambda_1\lambda_5k_1}^{{\bf
      q}} X_{\lambda_1\lambda_5k_2}^{{\bf
      q}\ast}\delta_{k_1,k_2}\delta_{\lambda_1,\lambda_3}\delta_{\lambda_2,\lambda_4}
  C_{{\bf q}\eta}(\Delta_{\lambda_1\lambda_5k_2})\nonumber\\
&&\mbox{} -X_{\lambda_4\lambda_2k_1}^{{\bf
 q}}X_{\lambda_3\lambda_1k_2}^{{\bf
 q}\ast} \delta_{\varepsilon_{\lambda_4}-\varepsilon_{\lambda_2}
-\varepsilon_{\lambda_3}+\varepsilon_{\lambda_1},(k_2-k_1)\Omega}
  C_{{\bf q}\eta}(\Delta_{\lambda_3\lambda_1k_2})]\Big\} +
\Big\{\lambda_1\leftrightarrow\lambda_2,\lambda_3\leftrightarrow\lambda_4\Big\}^{\ast}
\label{Lambda}
\end{eqnarray}
\end{widetext}
being a time-independent tensor in the case without
degeneracy. $\{\lambda_1\leftrightarrow\lambda_2,\lambda_3\leftrightarrow\lambda_4\}^{\ast}$
in the above equation stands for the same terms as in the previous
$\{\}$ but interchanging $\lambda_1$ and $\lambda_2$, $\lambda_3$ and
$\lambda_4$ and taking a complex conjugate.
In the zero driving field limit, the Floquet wavefunction reduces from
a many-frequency one to a single-frequency one.
Thus, the only nonzero contribution in the summation
$X_{\lambda_1\lambda_2}^{{\bf q}} = \sum_{k}
e^{i\Delta_{\lambda_1\lambda_2k}t}X_{\lambda_1\lambda_2k}^{{\bf q}}$
is the term with $k=k_0$, with
$\hbar\Delta_{\lambda_1\lambda_2k_0}=E_{\lambda_1}-E_{\lambda_2}$ being
the energy difference between states $|\lambda_1\rangle$ and
$|\lambda_2\rangle$.
 $\delta_{\varepsilon_{\lambda_4}-\varepsilon_{\lambda_2}
-\varepsilon_{\lambda_3}+\varepsilon_{\lambda_1}, (k_2-k_1)\Omega}$
can be simplified into $(\delta_{\lambda_1,\lambda_3}
\delta_{\lambda_2,\lambda_4}+\delta_{\lambda_1,\lambda_2}
\delta_{\lambda_3,\lambda_4})$.
Therefore Eq.\ (\ref{steady}) can be simplified
  into
\begin{eqnarray}
\frac{\partial}{\partial
  t}\rho_{\lambda_1\lambda_2}^e&=&\Big\{-\frac{\pi}{\hbar^2} \sum_{{\bf
    q}\eta,\lambda_3}|M_{{\bf q}\eta}|^{2}|\langle\lambda_1|e^{i{\bf
      q}\cdot{\bf r}}|\lambda_3\rangle|^2\
\rho_{\lambda_1\lambda_2}^e\nonumber\\
&&\mbox{}\times[\bar{n}(\omega_{{\bf
      q}\eta})
\delta(E_{\lambda_1}-E_{\lambda_3}+
   \omega_{{\bf q}\eta})\nonumber\\
&&+(\bar{n}(\omega_{{\bf
       q}\eta})+1)
   \delta(E_{\lambda_1}-E_{\lambda_3}-\omega_{{\bf q}\eta})]\Big\}\nonumber\\
&& + \Big\{\lambda_1\leftrightarrow\lambda_2\Big\}^{\ast}
\end{eqnarray}
for $\lambda_1\ne \lambda_2$, where $E_{\lambda}$ is the energy of the state $|\lambda\rangle$ and
\begin{eqnarray}
\frac{\partial}{\partial t} \rho_{\lambda_1\lambda_1}^e&=&
 -\frac{2\pi}{\hbar^2} \sum_{{\bf q}\eta}
  |M_{{\bf q}\eta}|^{2} \Big\{\sum_{\lambda_3}|
\langle\lambda_1|e^{i{\bf q}\cdot{\bf
      r}}|\lambda_3\rangle|^2\rho_{\lambda_1\lambda_1}^e\nonumber\\
&&\mbox{}\times [\bar{n}(\omega_{{\bf
    q}\eta})
\delta(E_{\lambda_1}-E_{\lambda_3}+\omega_{{\bf q}\eta})
\nonumber\\
&&\mbox{}+(\bar{n}(\omega_{{\bf
    q}\eta})+1)\delta(E_{\lambda_1}-E_{\lambda_3}-\omega_{{\bf q}\eta})\ ]\nonumber\\
&&\hspace{-2.1cm}-\sum_{\lambda_2}|\langle\lambda_1|e^{i{\bf q}\cdot{\bf
    r}}|\lambda_2\rangle|^2\rho_{\lambda_2\lambda_2}^e\ [\bar{n}(\omega_{{\bf
    q}\eta})\delta(E_{\lambda_2}-E_{\lambda_1}+\omega_{{\bf q}\eta})\nonumber\\
&&\hspace{-0.9cm}+(\bar{n}(\omega_{{\bf
    q}\eta})+1)\delta(E_{\lambda_2}-E_{\lambda_1}-\omega_{{\bf q}\eta})\ ]\Big\}\ .
\end{eqnarray}
These equations are consistent with the
kinetic spin Bloch equations.\cite{Weng}

Equation (\ref{steady}) can be rewritten in the matrix form as
$\frac{\partial}{\partial t}{\bf \rho}^e= - {\bf \Lambda}{\bf \rho}^e$,
which is a standard first order differential equation. It
can be solved through the eigenvalues and eigenvectors of the
matrix ${\bf \Lambda}$. Thus, for any observable ${\hat O}$,
\begin{eqnarray}
O(t) &=& \mbox{Tr}({\hat O}{\bf {\rho}}^e) \nonumber\\
  &=&\sum_{\lambda_1\cdots\lambda_6}
  \langle\lambda_2(t)|{\hat O}|\lambda_1(t)\rangle
  {\bf P}_{(\lambda_1\lambda_2)(\lambda_3\lambda_4)}\nonumber\\
&&\mbox{}\times e^{-{\bf
 \Gamma}_{(\lambda_3\lambda_4)}t} {\bf
  P}_{(\lambda_3\lambda_4)(\lambda_5\lambda_6)}^{-1}
\rho_{\lambda_5\lambda_6}^e(0)
\label{Oeq}
\end{eqnarray}
with ${\bf \Gamma}={\bf P}^{-1}{\bf \Lambda}{\bf P}$ being a diagonal
matrix and ${\bf P}$ being the transformation matrix. By
 solving Eqs.\ (\ref{psi}), (\ref{eigen}),
 (\ref{Lambda}) and (\ref{Oeq}) numerically with an initial
spin polarization $S_z(0)$, one obtains the
time evolution of $S_z$.

\section{Numerical results}
We consider an isolated 1D InAs QD, where the low-lying states can be
approximated by eigenstates in an infinite-well-depth
potential:\cite{1DQD2} $V_c({\bf r})=0$ if
$0<x<L_a$, $0<y<L_b$ and $0<z<L_c$ and $V_c({\bf r})=\infty$ elsewhere.
We choose $L_a = L_b = 20$\ nm and $L_c = 70$\ nm in the calculation.
The separation between the first
and the second subbands is about 9\ meV (15\ THz) along the
$z$-direction, and 120\ meV along the $x$ (or $y$)-direction.
By averaging over the lowest states in the $x$ and $y$ directions,
one can turn the problem into an effective 1D problem. We apply a static
magnetic field $B_0$ of 0.5/0.7/1\ T along the $z$-axis which corresponds to a
Zeeman splitting of about 0.4/0.56/0.8\ meV (0.62/0.87/1.25\ THz).
In this energy range the electron-acoustic-phonon scattering is dominated by
the deformation potential coupling in InAs. The corresponding
scattering matrix reads
$M_{{\bf q}{sl}} = \Xi \sqrt{\hbar q/2Dv_{sl}}$, where $\Xi$ is the
deformation potential, $D$ denotes the volume density and $v_{sl}$ stands for
the longitudinal sound velocity. All the parameters used in
the calculation are listed in Table\ I.\cite{para}
We take the cut-off frequency
of the phonon reservoir to be the Debye frequency $\omega_{D}$.
The temperature is taken to be 200\ mK,
corresponding to an energy of 0.016\ meV (0.026\ THz), which is quite
smaller than the other energies of the system (especially the Zeeman
splitting energy), {\em i.e.}, we study the spin relaxation in low
temperature regime where the phonon-absorption processes are
energetically unfavorable.
The Rashba parameter is taken to be $\alpha_R=3.0\times10^{-9}$\
eV$\cdot$cm.\cite{Grundler}
By including all the scattering processes between the Floquet states
due to the electron-phonon scattering, one can calculate
the scattering matrix ${\bf \Lambda}$ [Eq.\ (\ref{Lambda})].
With a preparation of occupying the first excited Zeeman state of
$H_0$ [see Eq.(4)] as the initial state,  one can obtain the time-evolution of
${S}_z$. By taking the envelope of $S_z$ and subtracting the equillibrium spin
polarization, we define $T_1$ as the time needed for decay of
the spin polarization by a factor of $1/e$.

\begin{table}[htbp]
\begin{tabular}{lllllll}
\hline\hline
D&\mbox{}&$5.9\times10^3$kg/m$^3$&\mbox{}\mbox{}&$g$&\mbox{}&$-14.7$\\
$v_{sl}$&\mbox{}&$4.28\times10^3$\ m/s&\mbox{}\mbox{}&$\omega_{D}$&\mbox{}&32.7\ THz\\
$\varXi$&\mbox{}&5.8\ eV&\mbox{}\mbox{}&$m^{\ast}$&\mbox{}&$0.0239$~$m_0$\mbox{}\\\hline\hline
\end{tabular}
\caption{Parameters used in the calculation}
\label{tab:parameter}
\end{table}

At the very low temperature we study, the spin relaxation is due to the
spin-flip transition between the lowest Zeeman sublevels.
Recently Fonseca-Romero {\em et al.} studied a model two-level
system coupled to  an Ohmic reservoir via $\sigma_x$.\cite{Fonseca-Romero}
They showed that the pseudo-spin relaxation and dephasing can be greatly
modified by the driving field when it is in the type of
$A\sigma_z\cos(\Omega t)$. Their results show that
at low temperature when the  frequency is below the cut-off
frequency of the reservoir, the driving field enhances the pseudo-spin
relaxation, otherwise impedes it.
However, in QDs, spin is coupled indirectly with the phonon bath via the
SOC. The effective spectral density of the spin-phonon coupling is
generally not Ohmic.\cite{Westfahl} Remarkably, in QD
this spectral density can be controlled by the QD geometry
(size, growth-direction, etc.), magnetic field, gate-voltage and the
strength and symmetry of the
SOC.\cite{Westfahl,Cheng2,Aleiner,dephasing,Sarma,Golovach2,Wang,Fabian,Falko}

\begin{figure}[thb]
\centerline{\psfig{figure=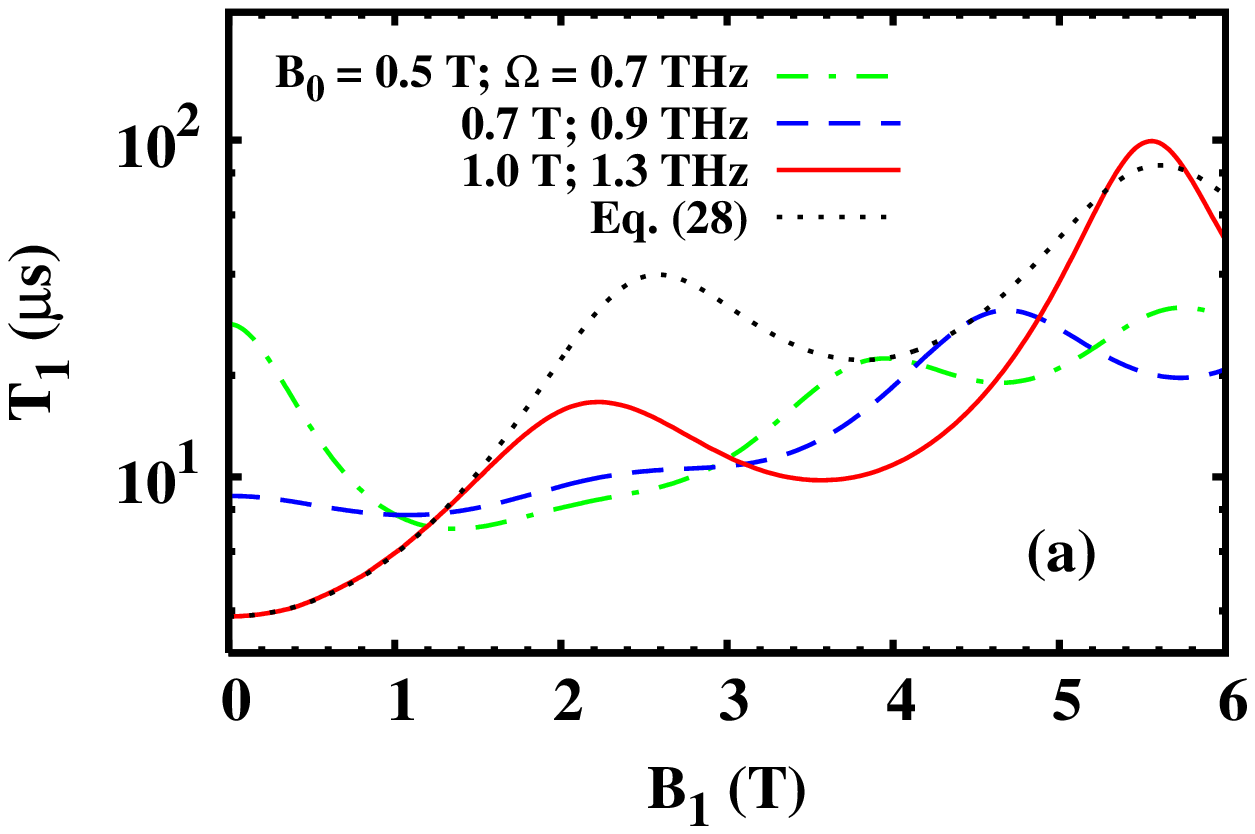,width=0.755\columnwidth}}
 \centerline{\psfig{figure=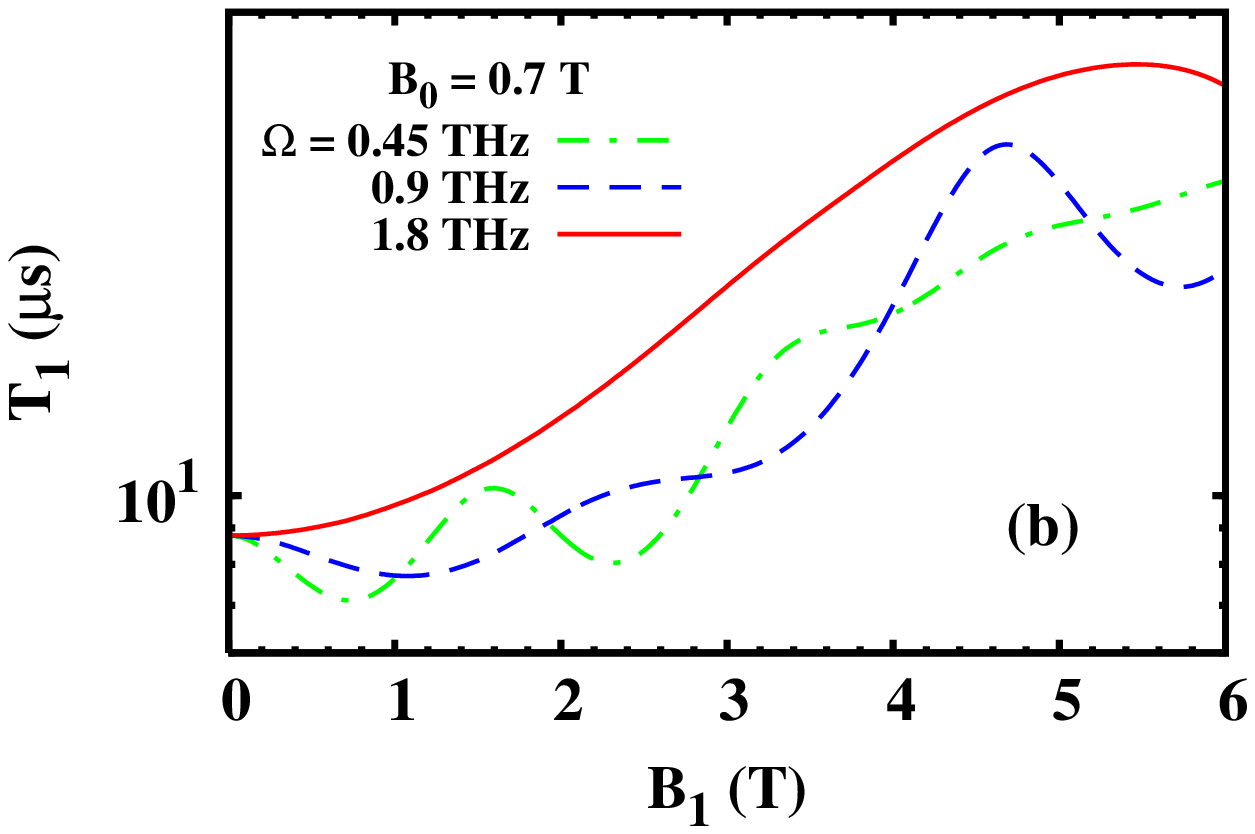,width=0.755\columnwidth}}
  \centerline{\psfig{figure=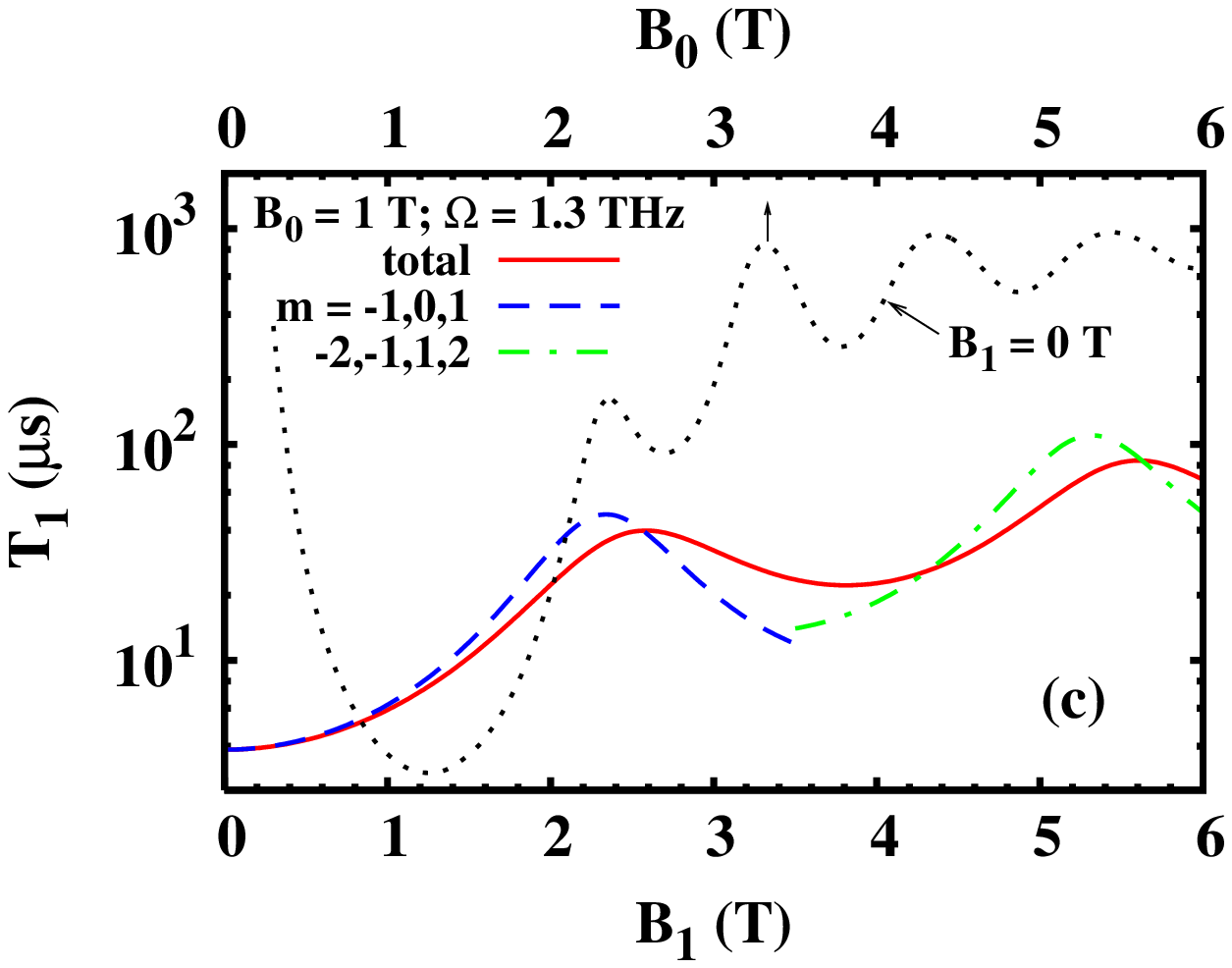,width=0.755\columnwidth}}
  \caption{(Color online) Spin relaxation time $T_1$ as a function of
    THz magnetic field strength $B_1$ for (a) $B_0=0.5$\ T and
    $\Omega=0.7$\ THz (green chain curve); $B_0=0.7$\ T
and $\Omega=0.9$\ THz (blue dashed curve); $B_0=1.0$\ T and
$\Omega=1.3$\ THz (red solid curve), all in resonance, and (b)
$B_0=0.7$\ T, $\Omega=0.45$\ THz (green chain curve); $0.9$\ THz
(blue dashed curve, in resonance);
    and $1.8$\ THz (red solid curve).
The black dotted curve in (a) is the spin relaxation time calculated from
the simplified model Eq.\ (\ref{approx}) for $B_0=1.0$\ T and
    $\Omega=1.3$\ THz.
(c) The spin relaxation time as a function
of $B_0$ without the THz magnetic field (black dotted curve),
and results from Eq.\
    (\ref{approx}) including all sidebands (red solid curve); only the
 0, and $\pm1$-th sideband (blue dashed curve), and only the $\pm1$ and $\pm2$-th sideband
 (green chain curve). Note the scale of $B_0$ is on the upper frame of the figure.}
\label{fig:THzB}
\end{figure}

In our study, we investigate the spin relaxation in an InAs 1D QD in
the presence of a THz driving magnetic field (no THz electric field included)
along the $z$-axis (whose
frequency is {\em below} the cut-off frequency) from a full
microscopic approach. By including enough base-states and using
the exact numerical diagonalization method,\cite{Cheng2} we can calculate
the Floquet wavefunctions. We then substitute these wavefunctions into
the kinetic equations Eq.\ (\ref{steady})
to obtain the dissipative dynamics of spin in QD
under the driving field. We find that differing form the
Ohmic spin-phonon coupling,\cite{Fonseca-Romero} the
spin relaxation time here can be either prolonged or
shortened by the sideband effect, depending on the frequency and
strength of both the THz
magnetic field and the  static magnetic field, with the latter providing
the Zeeman splitting.

We plot the spin relaxation time $T_1$ as a function of the THz magnetic field
$B_1$ in Fig.\ 1(a) with the electric field $E=0$. The green-chain/blue-dashed/red-solid
curves correspond to $B_0=$0.5/0.7/1.0\ T and $\Omega=$0.7/0.9/1.3\ THz
respectively. The THz magnetic field is tuned to be in resonance with  the
Zeeman splitting. It is seen from the figure that the spin relaxation time
$T_1$ increases with the strength of the THz magnetic field $B_1$,
with some modulations.  It increases more rapidly for the
case with $B_0$ = 1.0\ T than the other two cases:
at $B_1\approx$ 5.6\ T, $T_1$ is increased about 25 times of the value
at $B_1=0$.  It is also noted that due to the different modulations,
when $B_1\le1$\ T,
the spin relaxation time $T_1$ decreases rapidly/mildly
for $B_0=0.5$\ T/0.7\ T
but increases rapidly for $B_0=1.0$\ T.
We also plot $T_1$ at $E=0$, and  $B_0=0.7$\ T as a function of $B_1$
with different  frequencies (0.45, 0.9, and 1.8\ THz) in Fig.\ 1(b).
Again the spin relaxation time increases overall with $B_1$  but
with different modulations. Nevertheless as a result of the modulation,
when $B_1\le0.7$\ T, $T_1$ decreases for the case with
$\Omega$=0.45\ THz and 0.9\ THz (mildly) but increases for the
case with $\Omega$=1.8\ THz.

These numerical results can be understood from the following simplified
analysis. At zero  electric field $E=0$, the effective Hamiltonian
[Eqs.\ (3-6)] in the matrix element form reads
\begin{eqnarray}
\langle n,\sigma|H_{eff}|n^{\prime},\sigma^{\prime}\rangle &=&
\Big\{E_n^0+\frac{1}{2}[E_z^0+E_z^1\cos(\Omega t)] \sigma\Big\}
\delta_{n,n^{\prime}}\nonumber\\ &&\hspace{-2.7cm}\times
\delta_{\sigma,\sigma^{\prime}}
+ E^{so}(n,\sigma;n^{\prime},\sigma^{\prime})\delta_{(-1)^{n+n^{\prime}}+1=0}
\delta_{\sigma+\sigma^{\prime}=0},
\label{eff1}
\end{eqnarray}
with $\langle z |n\rangle = \sqrt{\frac{2}{L_c}}\sin(\frac{n\pi
  z}{L_c})$, and $|\sigma\rangle$, the eigen vector of $\sigma_z$.
$E_n^0=n^2 {\hbar^2\pi^2}/({2m^{\ast}L_c^2})$ is the subband energy.
 $E_z^0=g\mu_B B_0$ and
$E_z^1=g\mu_B B_1$ represent the spin splittings due to the static and
the THz magnetic field. $E^{so}(n,\sigma;n^{\prime},-\sigma) = \sigma
\alpha_R \langle n|\frac{-i}{\hbar}p_z| n^{\prime} \rangle =
\sigma\frac{4 \alpha_R n n^{\prime}}{L_c(n^{\prime 2}-n^2)}$.
The last term of Eq.\ (\ref{eff1}), which is the energy due to the SOC,
is nonzero only for
states with different spin and different parity.
For very low temperature  and for not too strong THz
magnetic field, we can restrict ourselves to consider only the lowest
two states due to the Zeeman splitting described by
the time-independent parts of the
effective Hamiltonian Eq.\ (\ref{eff1}).
These two states, denoted as $|\xi_{\sigma}\rangle$ with
$\sigma=\uparrow,\downarrow$ ($+,-$),
read
\begin{equation}
|\xi_{\sigma}\rangle \approx |1,\sigma\rangle
+\sum_n\frac{E^{so}({2n,-\sigma;1,\sigma})}{E_1-E_{2n}+E_z\sigma}|2n,-\sigma\rangle\ ,
\label{xi}
\end{equation}
from the perturbation.
The matrix elements of the effective Hamiltonian
in the space of these two states are
  $\langle \xi_{\sigma}|H_{eff}|\xi_{\sigma^{\prime}}\rangle \approx
  \frac{1}{2}[E_z+g\mu_{B} B_1\cos(\Omega t)]
  \sigma\delta_{\sigma,\sigma^{\prime}}$
approximately by retaining only the dominant terms.
$E_z$ is the energy difference
of the lowest two levels of $H_0$ which is approximately $g\mu_{B} B_0$
for not too large $B_0$.
For this Hamiltonian the Sch\"odinger equation can be integrated out
directly. The Floquet wavefunctions are therefore
\begin{eqnarray}
  \hspace{-0.45cm}\Psi_{\sigma}(t)&&\hspace{-0.2cm}=
  \exp\{-i[\frac{\sigma g\mu_{B} B_0 t}{2 \hbar}  +
    \frac{\sigma g\mu_{B} B_1}{2\hbar\Omega}\sin(\Omega t)]\} \xi_\sigma
\nonumber\\
  &&\hspace{-0.2cm}= \exp(-i\frac{\sigma g\mu_{B} B_0 t}{2\hbar}) \sum_m
J_m(\frac{\sigma g\mu_{B} B_1}{2\hbar\Omega})\nonumber\\
  &&\times \exp(-im\Omega t)\xi_\sigma\ ,
\label{waf}
\end{eqnarray}
in which ${J}_m$ is the $m$-th Bessel function of the first kind and
$\xi_\sigma$ represents the pseudo-spin wavefunction of the lowest two
states [Eq.\ (\ref{xi})]. The form of the wavefunction clearly indicates
the sideband effect. Specifically, the probability of finding the electron in
$|\Psi_{\sigma}\rangle$   is  $|{J}_m(\frac{\sigma
  g\mu_{B} B_1}{2\hbar\Omega})|^2$ which oscillates at a frequency $\sigma
g\mu_{B} B_0/\hbar + m\Omega$. The frequency together with
its corresponding coefficient $J_m(\frac{\sigma g\mu_{B}
  B_1}{2\hbar\Omega})$ is referred to as the  $m$-th sideband from
  now on. In the following we show that this sideband effect
can greatly affect the spin relaxation.

Substituting these wavefunctions into the kinetic equation Eq.\ (\ref{EOM}),
one obtains the following scattering matrix
\begin{eqnarray}
 \hspace{-0.4cm} \Lambda_{\uparrow\uparrow\uparrow\uparrow}
&=& \frac{2\pi}{\hbar^2}
  \sum_{\mathbf{q}\eta} |M_{\mathbf{q}\eta}|^2 \sum_k
  |X^{\mathbf{q}}_{\uparrow\downarrow k}|^2
  C_{\mathbf{q}\eta}(\Delta_{\uparrow\downarrow k})\nonumber \\
  &&\hspace{-1cm} = \sum_k \Gamma(\Delta_{\uparrow\downarrow k})
  \sum_{m,m^{\prime}}{J}_{-m}{J}_{m+k}{J}_{-m^{\prime}}
{J}_{m^{\prime}+k}(\frac{g\mu_{B} B_1}{2\hbar\Omega}) \nonumber\\
&&\hspace{-1cm}= \sum_k \Gamma(\Delta_{\uparrow\downarrow k})
{J}_k^2(\frac{g\mu_{B} B_1}{\hbar\Omega})\ .
\label{LL}
\end{eqnarray}
Here the relation $\sum_{m}
{J}_{-m}{J}_{m+k}(x) = {J}_{k}(2x)$ is applied and
\begin{eqnarray}
&&\hspace{-0.5cm}\Gamma(\Delta_{\uparrow\downarrow k})=\frac{2\pi}{\hbar^2}
\sum_{\mathbf{q}\eta} |M_{\mathbf{q}\eta}|^2
|\langle\xi_\uparrow|e^{i\mathbf{q}\cdot\mathbf{r}}|\xi_\downarrow
\rangle|^2[\bar{n}(\omega_{{\bf
    q}\eta})\nonumber\\
&&\hspace{-0.5cm}\times\delta(\Delta_{\uparrow\downarrow k}+\omega_{{\bf
    q}\eta})+(\bar{n}(\omega_{{\bf
    q}\eta})+1)\delta(\Delta_{\uparrow\downarrow k}-\omega_{{\bf
    q}\eta})]\ .
\end{eqnarray}
 Without the driving field, Eq.\ (\ref{LL}) reduces back to the well known form
$\Lambda_{\uparrow\uparrow\uparrow\uparrow}=\Gamma(\Delta_{\uparrow\downarrow 0})$.
It is noted that unlike the driving-field free case where only phonons with energy
$g\mu_{B} B_0$ contribute to the spin relaxation,
due to the sideband effect caused by the driving field, from Eq.\ (\ref{LL})
one can see that phonons with energies $\Delta_{\uparrow\downarrow k} = g\mu_{B} B_0
+ k\hbar\Omega$ ($k\ne0$) also contribute to the spin
  relaxation.
The spin relaxation rate is $T_1^{-1}=2\Lambda_{\uparrow\uparrow\uparrow\uparrow}$.
Particularly, at zero driving field, $T_1^{-1} =
2\Gamma(g\mu_{B} B_0)$. The spin relaxation time as a
function of the THz magnetic field $B_1$ and the static magnetic field $B_0$,
$T_1(B_1,B_0)$, reads
\begin{eqnarray}
 &&\hspace{-1cm}T_1(B_1,B_0)\approx \
T_1(0,B_0)/\Big\{{J}_0^2(g\mu_{B} B_1/\hbar\Omega) \nonumber\\
&&\hspace{-0.7cm}+(\sum_{k=-\infty}^{-1}+\sum_{k=1}^{\infty})
{J}_k^2(\frac{g\mu_{B} B_1}{\hbar\Omega})
  \frac{T_1(0,B_0)}{T_{0}(g\mu_{B} B_0+k\hbar\Omega)}\Big\}
\label{approx}
\end{eqnarray}
in which $T_{0}(g\mu_{B} B_0+k\hbar\Omega)$ is the spin relaxation time at
same external condition (QD geometry, static magnetic field,
temperature, {\em etc.}) but at different energy
$g\mu_{B} B_0+k\hbar\Omega$. We approximately take
$T_{0}(g\mu_{B} B_0+k\hbar\Omega)$ as the spin relaxation time
$T_1(0, B_0^{\ast})$, where $B_0^{\ast}$ is determine by the condition
that the corresponding  Zeeman splitting for the lowest two states
is $g\mu_{B} B_0+k\hbar\Omega$.
This approximation is reasonable when $g\mu_{B} B_0+k\hbar\Omega$ is much
smaller than the energy difference between the first and second
subbands, since the difference in the spin mixing
for the lowest two states between the case with a static magnetic field
$B_0^{\ast}$ and the case with $B_0$ is marginal.

We first calculate $T_1(0,B_0)$ as a function of $B_0$ and plot it
in Fig.\ 1(c) as black dotted curve. Facilitated with this quantity,
we further obtain $T_1(B_1,B_0)$ in Eq.\ (\ref{approx}) as a
function of $B_1$ for $B_0=1$\ T and $\Omega=1.3$\ THz and plot it
as black dotted curve in Fig.\ 1(a). It is seen that this
approximate results agrees  with the numerical result qualitatively.
It is seen from the simplified model that the dominant contribution
comes from the term of ${J}_0^2(\frac{\sigma g\mu_{B}
B_1}{\hbar\Omega})$ in the denominator in Eq.\ (\ref{approx}) which
is zero at $B_1\approx$2.4 and 5.6\ T, corresponding to the first
and the second peaks in Fig.\ 1(a). In order to reveal which
sideband contributes  to the peaks, we plot $T_1(B_1,B_0)$
calculated from Eq.\ (\ref{approx}) with all the sidebands [red
solid curve, same as the black dotted curve in Fig.\ 1(a)], with
only the 0, $\pm1$-th sidebands (blue dashed curve), and with
only the  $\pm1$, $\pm2$-th sidebands (green chain curve) in
Fig.\ 1(c) versus $B_1$ when $B_0$=1\ T and $\Omega=1.3$\ THz. It is
seen in the figure that the result with only the  0, $\pm1$-th
sideband agrees with the total approximation result pretty well when
$B_1< 3.5$\ T which indicates that the first peak is mainly due to
these sidebands. When $B_1>3.5$\ T, the result including only the
$\pm1$, and $\pm2$-th sidebands is in reasonable agreement  with the
result with all the sidebands. Therefore the second peak mainly
comes from the $\pm1$ and $\pm2$-th sidebands. The oscillations in
spin relaxation time are mainly due to oscillations of the sideband
amplitude with the strength of the driving field, {\em i.e.}, the
sideband factor ${J}_k^2(g\mu_{B} B_1/\hbar\Omega)$ in Eq.\
(\ref{approx}). The relaxation time $T_{0}(g\mu_{B}
B_0+k\hbar\Omega)$ can be larger or smaller than $T_1(0,B_0)$
depending on the details of the SOC mediated spin-phonon
coupling,\cite{Westfahl,Cheng2,Aleiner,dephasing,Sarma,Golovach2,Wang,Fabian}
which determines the effect of the THz magnetic field on the spin
relaxation time.

At very low temperature, if $g\mu_{B} B_0+k\hbar\Omega$ is less than
zero, the relaxation process is prohibited because there is no phonon
to be absorbed. Therefore, $T_1(0,B_0)/T_{0}(g\mu_{B} B_0+k\hbar\Omega)$ is
zero for $g\mu_{B} B_0 + k \hbar \Omega < 0$.
For the case of resonant driven system, only  $k\ge0$ terms
remain finite. Among these terms  only the terms with  $k=0$ and 1
are important for small $B_1$.
${J}_0^2(g\mu_{B} B_1/\hbar\Omega)\approx
1-2{J}_1^2(g\mu_{B} B_1/\hbar\Omega)$ if $g\mu_{B}
B_1/\hbar\Omega<1$. Thus the denominator is
approximatly $1+{J}_1^2(g\mu_{B}
B_1/\hbar\Omega)[T_1(0,B_0)/T_{0}(g\mu_{B} B_0+\hbar\Omega)-2]$.
Therefore, $T_{0}(g\mu_{B} B_0+\hbar\Omega)$ becomes an
important factor for $T_1(B_1,B_0)$. The three color
curves in Fig.\ 1(a) with $B_0 =1.0$/0.7/0.5\ T,
corresponding to $T_1(B_1,B_0)$
being increased/insensitive/decreased with $B_1$ when
$B_1\le 1.1$\ T. This is because $T_{0}(g\mu_{B} B_0+\hbar\Omega)\approx
T_1(0,2B_0)$ is larger than/approximately/smaller than $T_1(0,B_0)/2$.
The same thing happens, if one changes the frequency of the THz
magnetic field, as $T_{0}(g\mu_{B} B_0+\hbar\Omega)$ also depends on
$\Omega$. The property of the frequency dependence in Fig.\ 1(b)
when $B_1<0.7$\ T can  be understood from the fact
that $T_{0}(g\mu_{B}
B_0+\hbar\Omega)\approx T_1(0,B_0+\hbar\Omega/g\mu_{B})$ is smaller/a
little smaller/larger than $T_1(0,B_0)/2$ when $\Omega=
0.45/0.9/1.8$\ THz.
This indicates that by properly tuning the frequency of
the THz magnetic field, we can change the effect of the THz magnetic
field efficiently.
For strong THz magnetic field, the the sideband factor
${J}_k^2(g\mu_{B} B_1/\hbar\Omega)$ is important only for terms with
large $k$ where $T_{0}(g\mu_{B} B_0+k\hbar\Omega)\approx
T_1(0,B_0+k\hbar\Omega/g\mu_{B})$ is  larger
than $T_1(0,B_0)$ for large enough $k$. Moreover, ${J}_k^2(g\mu_{B}
B_1/\hbar\Omega)$ also decreases with $B_1$. Therefore for strong
enough THz magnetic field, the  spin relaxation time is always larger
than $T_1(0,B_0)$ as indicated by the six colored curves in Figs.\ 1(a) and (b).

\section{Conclusions}
In conclusion, we apply the Floquet-Markov theory to the spin kinetics in
1D InAs QD to study the spin relaxation in the presence of the THz
driving field. Especially, we study that the spin relxation under a THz
magnetic field which is parallel to a static magnetic field. We
find that the spin relaxation time can be effectively manipulated by
the driving field depending on its frequency and strength.
This offers a new way to control the spin relaxation.
The effect is understood as the sideband effect modulates the indirect
spin-phonon coupling. The effect of the driving field also depends on the
properties of the QD, such
as the QD geometry, the strength and symmetries of the
spin-orbit coupling, etc., which can be tuned by the gate-voltage, and
the static magnetic field.
The formulism developed here can be generalized to other
systems, such as the two-dimensional electron/hole gas with the SOC
to study the spin relaxation. The corresponding kinetic
equation can further be used to study the problems such as
the AC-field-induced spin
polarization and the related spin transport. These  are
still under investigation and will be published elsewhere.

\begin{acknowledgments}
This work was supported by the Natural Science Foundation of China
  under Grant No. 10574120, the National Basic Research Program of China under
Grant No.\ 2006CNBOL1205,   the Natural Science Foundation
  of Anhui Province under Grant No. 050460203, the Innovation
  Project of Chinese Academy of Sciences and SRFDP.
One of the authors (JHJ) would like to thank J. L. Cheng for helpful discussions.
\end{acknowledgments}

\end{document}